\documentclass{iopart}
\usepackage{amssymb}
\usepackage{siunitx}  
\usepackage{graphicx}
\usepackage{hyperref}
\usepackage[utf8]{inputenc}
\usepackage[T1]{fontenc}
\usepackage{color}

\newcommand{\der}{\mathrm{d}}

\begin{document}

\title[Scattered light modeling and subtraction]{End benches scattered
    light modeling and subtraction in Advanced Virgo}

\newcommand*{\LAPP}{Laboratoire  d’Annecy  de  Physique  des
  Particules  (LAPP),  Univ. Grenoble  Alpes,
  Université  Savoie  Mont Blanc,
  CNRS/IN2P3,  F-74941  Annecy,  France
}

\author{M~Wąs, R~Gouaty, R~Bonnand} 

\address{\LAPP} 
\ead{michal.was@lapp.in2p3.fr}

\date{\today}

\begin{abstract}
Advanced Virgo end benches were a significant source of scattered
light noise during the third observing run that lasted from April 1
2019 until March 27 2020. We describe how that noise could be
subtracted using auxiliary channels during the online strain data
reconstruction. We model in detail the scattered light noise coupling
and demonstrate that further noise subtraction can be achieved. We
also show that the fitted model parameters can be used to optically
characterized the interferometer and in particular provide a novel way
of establishing an absolute calibration of the detector strain data. 
\end{abstract}

\pacs{
04.80.Nn, 95.55.Ym, 
42.25.Fx 
}

\maketitle

\section{Introduction}
Interferometric gravitational wave detectors have their sensitivity
affected by scattered light, especially when microseism ground motion
is elevated at times of rough seas. Examples of ground motion coupling
to the sensitivity of detectors through scattered light have been
previously described for initial Virgo~\cite{virgoDetchar,virgoScattered10},
GEO-HF~\cite{GEOHFprogram} and most recently advanced LIGO~\cite{O1_DQCBC,aLIGO_O3_scatter}.

In this paper we focus on the coupling of light scattered by the end
suspended benches to the sensitivity of Advanced Virgo. This was a
significant
source of scattered light noise during the third observing
run (O3) that lasted from April 1 2019 until March 27 2020. However, it was
sufficiently well measured that scattered light noise could be
subtracted after the fact from the gravitational wave strain data.

This paper is organized as follows. In section~\ref{sec:theory} we
describe the theory of scattered light coupling from the suspended
benches, in section~\ref{sec:measurement} we show that scattered light can
be measured using photodiodes located on these benches, in
section~\ref{sec:subtract} we demonstrate how these signals can be
used to subtract scattered light noise from the strain data, and in
section~\ref{sec:calibration} we propose how to use scattered light as
a new method to calibrate the strain data.

\section{End benches scattered light theory}
\label{sec:theory}

\begin{figure}
  \centering
  \includegraphics[width=0.7\textwidth]{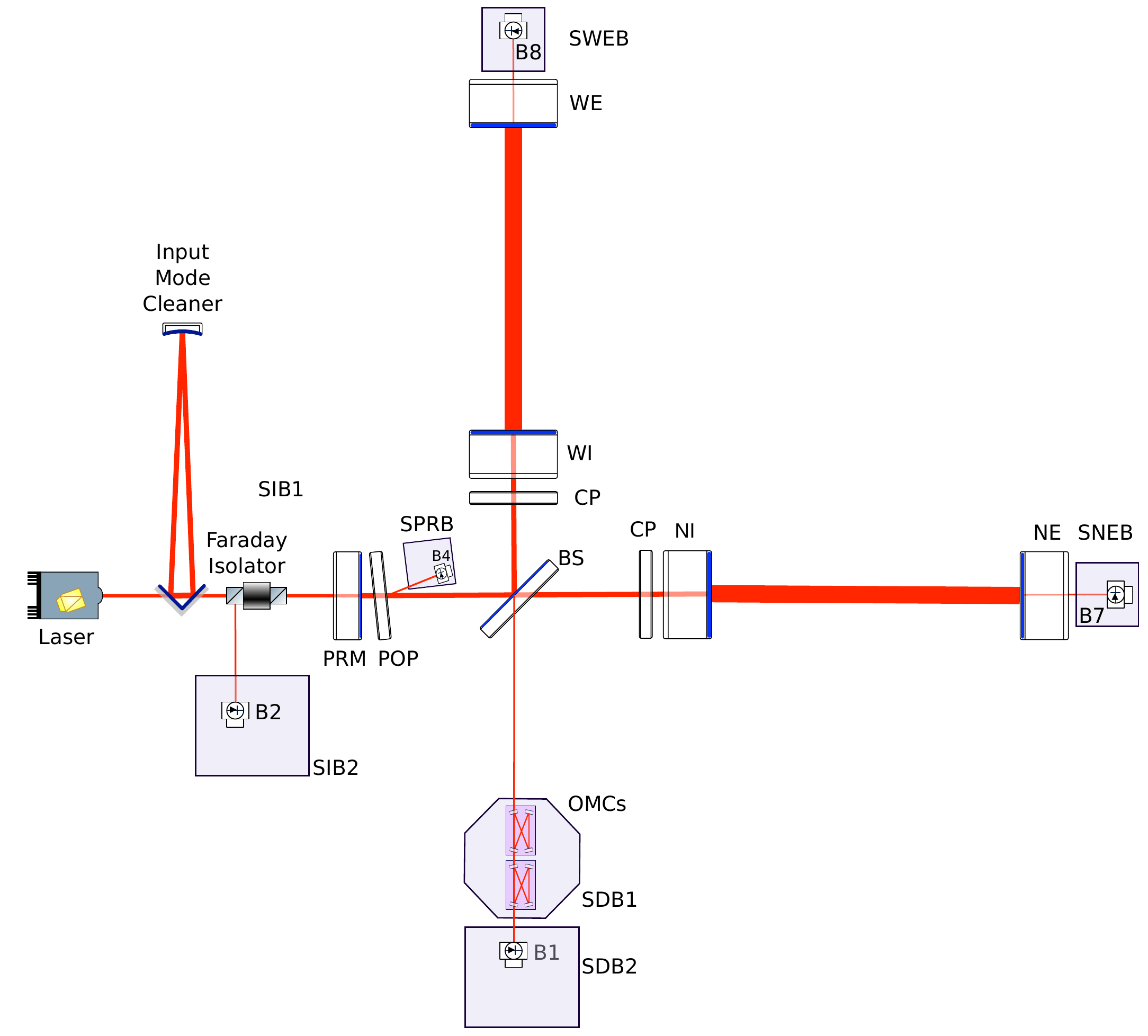}
  \caption{Optical layout of Advanced Virgo during O3, adapted from
    \cite{aVirgoStatus18} with kind permission of The European Physical Journal (EPJ).}
  \label{fig:layout}
\end{figure}

A simplified optical layout of Advanced Virgo during O3 is shown on
figure~\ref{fig:layout}. It is a power recycled Fabry-Perot Michelson
interferometer with 3\,km arms. The interferometer transforms
gravitational wave strain into audio frequency power fluctuations on
the anti-symmetric port photodiode denoted B1. In transmission of the
north end (NE) mirror and respectively the west end (WE) mirror are
located the suspended north end bench (SNEB) and respectively
the suspended west end bench (SWEB). Each of these benches host among
others a photodiode denoted respectively B7 and B8. 

Let $E_0$ be the electromagnetic field inside the west Fabry-Perot
cavity at the highly reflective (HR) surface of the WE mirror of power
transmission $T_\text{WE} \simeq 4.3\times 10^{-6}$~\cite{WEtransmission},
and $x$ be the distance between that surface and the scattering
surface located on SWEB which reflects a fraction $f_r$ of the
impinging light. The electromagnetic field inside the Fabry-Perot
cavity is in the fundamental Gaussian transverse mode, hence the field
forward propagating on the bench is also in this mode, in particular
light reaching photo-diodes. On the contrary light is scattered at all
angles, the fraction $f_r$ considers only the small portion of
scattered light that
is back scattered in the fundamental Gaussian mode and mode matched
with the light inside the Fabry-Perot cavity. This is the only
scattered light which will efficiently interfere with the dominant field in the
arm cavities and on the photodiodes.
As $T_\text{WE}$ and $f_r$ are small in the following
we will keep only the leading term in $\sqrt{T_\text{WE}}$ and $\sqrt{f_r}$.

The scattering surface create a field
$E_\text{sc}$ that at the HR surface of the WE mirror is
\begin{equation}
  E_\text{sc} = \sqrt{T_\text{WE} f_r} E_0 e^{ i \phi_\text{sc}}, 
\end{equation}
where $\phi_\text{sc} = 4 \pi \frac{x}{\lambda}$ is the phase delay due to the
round trip propagation and $\lambda$ is the laser wavelength. A small fraction of that
light is transmitted through the HR coating yielding a total field
\begin{equation}
  E_\text{tot} = E_0 + \sqrt{T_\text{WE}} E_\text{sc} = E_0 \left(1 + T_\text{WE} \sqrt{f_r} e^{ i \phi_\text{sc} }\right), 
\end{equation}
while the majority is reflected back
yielding a field $E_\text{B8}$ towards the B8 photodiode
\begin{equation}
  E_\text{B8} = \sqrt{T_\text{WE}} E_0 + E_\text{sc} = \sqrt{T_\text{WE}} E_0 \left(1 + \sqrt{f_r} e^{ i \phi_\text{sc}}\right). 
\end{equation}
The field perturbation inside the arm cavity is amplified by the
Fabry-Perot cavity in a frequency dependent way by
\begin{equation}
 G(f) =  \frac{1}{1-r} \frac{1}{1 + i \frac{f}{f_\text{arm}}},\label{eq:cavity_gain}
\end{equation}
where $r=\sqrt{1-T_\text{WI}-\Lambda_\text{arm}} \sim 0.993$~\cite{WItransmission} is the
effective field reflectivity of
the WI mirror taking into account the arm cavity round trip losses
$\Lambda_\text{arm}$,
\begin{equation}
f_\text{arm} = \frac{1-r}{r} \frac{c}{4\pi L}\simeq 55\,{\rm Hz} \label{eq:cavity_pole}  
\end{equation}
 is the arm cavity
pole frequency and $L=3$\,km is the arm cavity length.
This yields a field inside the cavity
\begin{equation}
  E_\text{cavity} = E_0 \left(1 +  G T_\text{WE} \sqrt{f_r} e^{ i \phi_\text{sc}}\right). 
\end{equation}

The interferometer is operated in DC read-out~\cite{Hild09, Fricke11}
with an offset in the differential arm length. Let us denote $\psi$
the differential phase offset and $h$ the amplitude of a putative
gravitational wave. The field inside the cavities becomes
\begin{eqnarray}
  E_\text{cavity, west} &=&  E_0  e^{ i \left(\frac{\psi}{2} + 2 \pi G \frac{h L}{\lambda} \right) } \left(1 +  G T_\text{WE} \sqrt{f_r} e^{ i \phi_\text{sc} }\right)\\
  E_\text{cavity, north} &=&  E_0  e^{ -i \left(\frac{\psi}{2} + 2 \pi G \frac{h L}{\lambda} \right) }
\end{eqnarray}
where $L$=3\,km is the cavity length. A fraction $T_\text{IM} =
0.0138$~\cite{WItransmission} of these fields returns through the input mirrors and
recombines at the beam splitter, which yield at the anti-symmetric
port of the interferometer a power
\begin{eqnarray*}
  P_\text{B1} &=& T_\text{IM} \left|E_\text{cavity,
                  west}-(1-\epsilon)E_\text{cavity, north}\right|^2\\
              &=& T_\text{IM} |E_0|^2 \left[
                  \psi^2 + \epsilon^2 + 2\psi 4 \pi G \frac{h
                  L}{\lambda} + \right.\\
  &&\left. 2\psi G T_\text{WE}
                  \sqrt{f_r}\sin\phi_\text{sc}
                  + 2\epsilon G T_\text{WE}
                  \sqrt{f_r}\cos\phi_\text{sc}
                  \right],
\end{eqnarray*}
where $\epsilon$ accounts for asymmetries between the two arms that
yield a contrast defect of the interferometer.

Thus scattered light directly mimicks a gravitational wave signal
through phase and amplitude coupling
\begin{eqnarray}
h_\text{sc, phase}  &=& \frac{1}{L}  T_\text{WE} \sqrt{f_r}
\frac{\lambda}{4 \pi}\sin \phi_\text{sc} \label{eq:sc_phase}\\
h_\text{sc, amplitude}  &=& \frac{\epsilon}{\psi}\frac{1}{L}  T_\text{WE} \sqrt{f_r}
\frac{\lambda}{4 \pi}\cos \phi_\text{sc}. \label{eq:sc_amplitude}
\end{eqnarray}

However, there are additional coupling path as scattered light
modulates the power inside the west arm cavity
\begin{equation}
  P_\text{cavity} = |E_0|^2  \left(1 + 2 G T_\text{WE} \sqrt{f_r} \cos
                    \phi_\text{sc}\right),
\end{equation}
which through radiation pressure displaces in opposite directions the
WI and WE mirrors and create an
additional spurious signal
\begin{eqnarray}
h_\text{sc, pressure}  &=& \frac{2}{L}  \frac{2 \delta P_\text{cavity}}{c}
                           \text{WE}_{F\rightarrow z} \\
                       &=& \frac{G}{L} \text{WE}_{F\rightarrow z}
                           \frac{8 |E_0|^2}{c}  T_\text{WE} \sqrt{f_r} \cos                                                                                                                                                                
                    \phi_\text{sc},\label{eq:sc_pressure}
\end{eqnarray}
where $c$ is the speed of light and
\begin{equation}
  \text{WE}_{F\rightarrow z}(f) =
  \frac{1}{M}\frac{1}{\Omega^2 - (2 \pi f)^2}  
\end{equation}
is the mechanical response of the suspended WE mirror with $M$ the
mass of the mirror and $\Omega$ the mirror suspension
pendulum angular frequency. Note that we omit here optical spring
effects that have a negligible effect above 10\,Hz as shown by the comparison to
numerical simulation described at the end of this section.

These power fluctuations are further amplified and filtered in the
combined power recycling and arm cavity by
\begin{equation}
 G_\text{combined}(f) =  \frac{\sqrt{T_\text{IM}}}{(1-r)(1-r_\text{PR}) + r_\text{PR}
  \Lambda_\text{arm}} \frac{1}{1 + i \frac{f}{f_\text{combined}}},
\end{equation}
where $r_\text{PR} = \sqrt{1-T_\text{PR}-\Lambda_\text{PR}} = 0.975$~\cite{PRtransmission} is
the effective field
reflectivity of the PR mirror taking into account the power recycling
cavity round trip losses $\Lambda_\text{PR}$ and
\begin{equation}
f_\text{combined} = \frac{(1-r)(1-r_\text{PR}) + r_\text{PR}
  \Lambda_\text{arm}}{r + r_\text{PR} - \Lambda_\text{arm}} \frac{c}{4\pi L}\simeq 1.0\,{\rm Hz} \label{eq:cavity_combined}  
\end{equation}
is the
combined cavity pole frequency. The displacement caused by the
radiation pressure of these fluctuations is common to
both arms, hence a priori the lengths of both arms is changed by the
same amount not yielding any differential signal. However the power in
the arms is not exactly equal, it may differ by a small fraction
$\rho$, which will yield a differential effect as radiation pressure
displacement is proportional to the power in the given arm. This
creates a spurious signal 
\begin{equation}
h_\text{sc, pressure recycled}  = \rho \sqrt{T_\text{WI}} \frac{G_\text{combined}}{2} h_\text{sc, pressure},\label{eq:sc_pressure_recycled}
\end{equation}
where the factor $\frac{1}{2}$ accounts for the recycled power being
split between the two arms.

\begin{figure}
  \centering
  \includegraphics[width=0.495\textwidth]{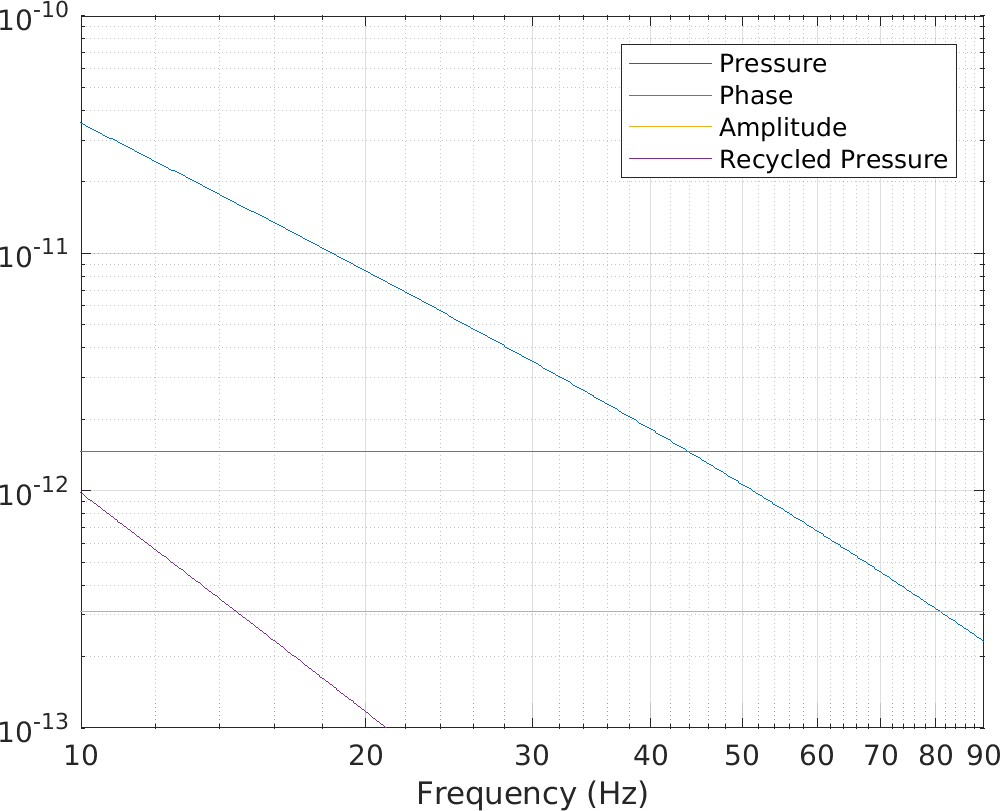}
  \includegraphics[width=0.495\textwidth]{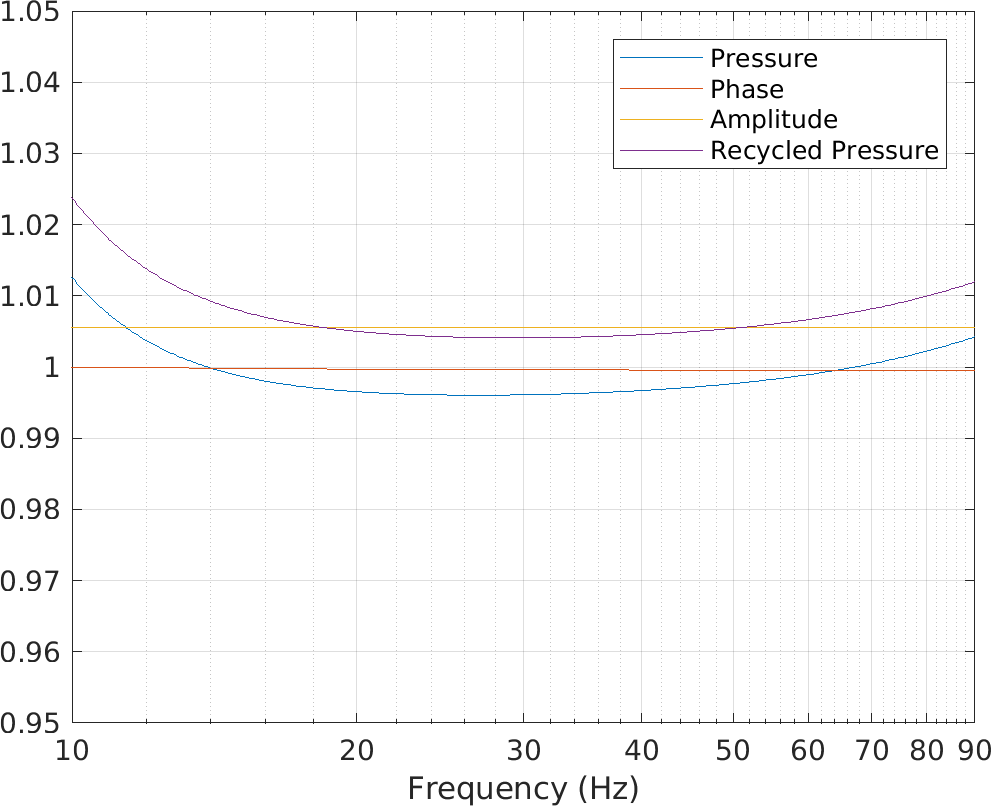}
  \caption{Left pane shows the scattering coupling transfer functions for the three
    coupling paths: radiation pressure given by equation \eref{eq:sc_pressure},
    phase given by equation \eref{eq:sc_phase}, amplitude given by
    equation \eref{eq:sc_amplitude}, and recycled radiation pressure
    given by equation \eref{eq:sc_pressure_recycled}. The right pane
    shows the ratio between the transfer function numerically computated
    using \emph{Optickle} and the analytical computation shown on the left pane.}
  \label{fig:TF}
\end{figure}

The power inside the arm cavities is estimated to be $|E_0|^2 = 90$\,kW
for a laser input power of 18\,W used in the first half of O3. While
the dark fringe power was set to be $P_\text{B1} = 2.8$\,mW and the
contrast defect light in the TEM00 mode in transmission of the output
mode cleaner was measured to be
$120\pm20$\,$\mu$W~\cite{contrastDefectO3a}, which yields
$\epsilon/\psi = 0.21$. The power ratio between the two arms has not
been accurately measured, but it varied by $\sim$1\% during the run
which sets an order of magnitude for $\rho$. These measurements allows us to evaluate the
relative contributions of the four coupling paths as shown on
figure~\ref{fig:TF} assuming a fiducial scattering $f_r =
10^{-6}$. The radiation pressure is dominant below 45\,Hz while the
phase coupling is dominant above. Note that in a previous estimate~\cite{Ottaway2012} of the
scattered light coupling a factor 2 was missing for the radiation
pressure and phase coupling path, and the amplitude and recycled
radiation pressure coupling paths were
completely neglected.

These analytic computations have been verified to be accurate with an
interferometer simulation performed with the \emph{Optickle}
simulation software \cite{optickle}. The only deviation is that in simulation the
radiation pressure effects include the optical spring at $\sim4$\,Hz,
which causes a small deviation at low frequency of 2\% in amplitude
and 1 degree in phase at $10$\,Hz as shown on figure~\ref{fig:TF}. In the
following section we use the optical simulation results scaled by the
parameters derived analytically to perform fits.

Fortunately the scattered light interference also produces a signal on the B8 photodiode
\begin{equation}\label{eq:P_B8}
  P_\text{B8} = \alpha T_\text{WE} |E_0|^2  \left(1 + 2 \sqrt{f_r} \cos
                    \phi_\text{sc} \right),
\end{equation}
where $\alpha$ accounts for the light losses between WE and the B8
photodiode due to pick-offs for quadrant photodiodes and cameras,
photodiode quantum efficiency, and other loss mechanisms. In
particular the relative intensity noise (RIN) of $P_\text{B8}$ yields
a direct measurement of the scattered light fraction
\begin{equation}\label{eq:P_B8_RIN}
  \frac{\delta P_\text{B8}}{P_\text{B8}} = 2 \sqrt{f_r} \cos
                    \phi_\text{sc}.
\end{equation}

\section{Measurement of scattered light fraction}
\label{sec:measurement}

\begin{figure}
  \centering
  \includegraphics[width=0.7\textwidth]{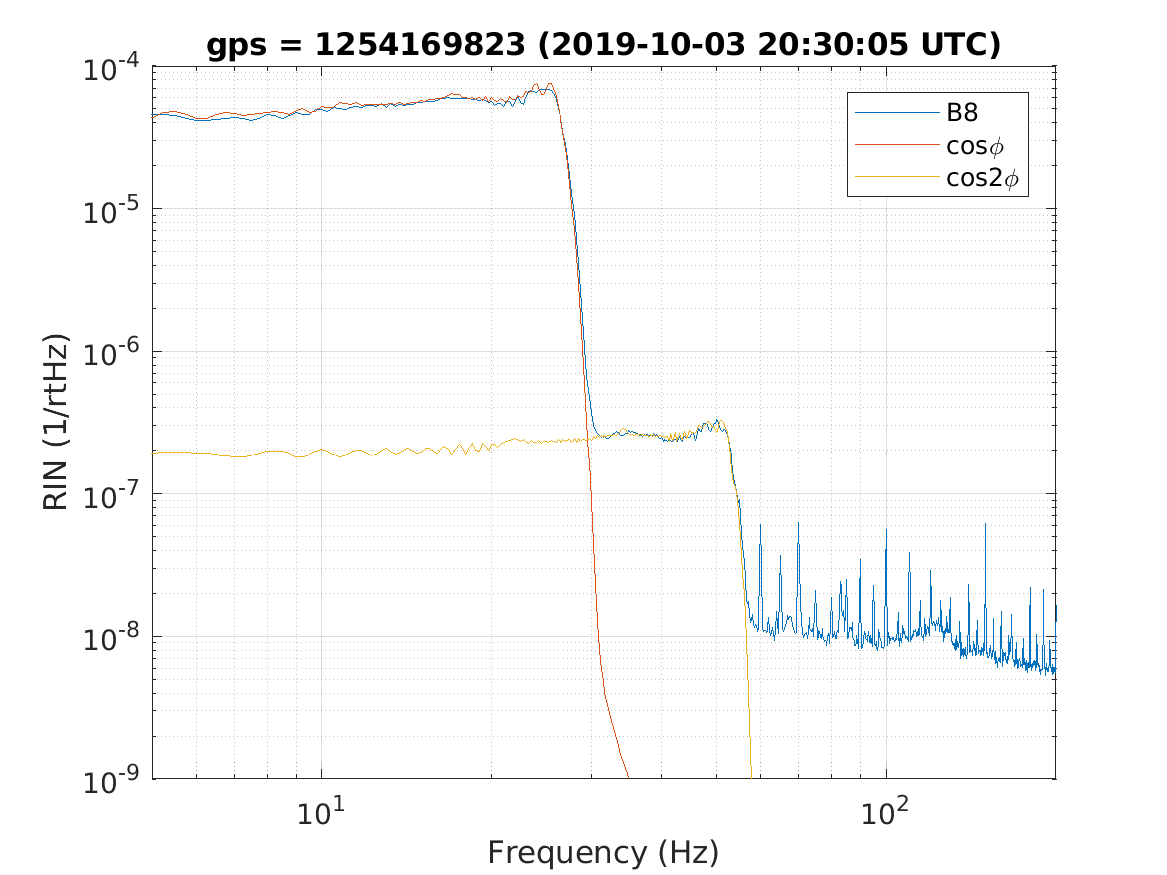}
  \caption{Relative intesity noise (RIN) of $P_\text{B8}$ with
    superposed fringe wrapped relative motion between SWEB and WE
    during the time of intentional SWEB motion with the quadrant
    shutters open.}
  \label{fig:fringeWrap}
\end{figure}

Large amplitude slow motion of the bench, especially at the microseism
peak at 300\,mHz, is up converted to the sensitive band of the
detector (above 10\,Hz) by the sine and cosine function of
$\phi_\text{sc} = 4\pi\frac{x}{\lambda}$. This yields a noise with characteristic
arch shape in a time-frequency representation of the data with a time
dependent frequency $f(t) = \frac{2\left|\dot{x}(t)\right|}{\lambda}$
proportional to the bench speed. The variation in distance
$x$ between SWEB and WE can be directly measured from the ground
connected local controls of SWEB (linear variable differential
transformers) and of WE (optical levers), where the
ground motion is removed at first order by taking the difference of
these sensors.

To study scattering coupling the SWEB bench has been intentionally
moved with large amplitude to increase the scattered light noise
signal in the detector. In total four measurements of 3 minutes in
duration have been performed over a 30 minutes time span. Two with an intentional motion of SWEB
and two with an intentional motion of SNEB. In each case one of the
measurement was with the quadrant shutters open and one with the
quadrant shutters closed.

\begin{table}
  \centering
  \begin{tabular}{l|cc}
    & $f_r$ & $f_{2r}$ \\
    \hline
    SWEB, open & $4.0 \times 10^{-8}$ & $1.5 \times 10^{-12}$ \\
    SWEB, closed & $ 6.0 \times 10^{-9}$ & $1.5 \times 10^{-12}$\\\hline 
    SNEB, open & $2.5 \times 10^{-8}$ & $3.0 \times 10^{-14}$ \\ 
    SNEB, closed & $3.5 \times 10^{-8}$ & $5.0 \times 10^{-14}$ 
  \end{tabular}
  \caption{Fitted scattered power for four different
    measurements with intentional large amplitude motion of either
    SWEB or SNEB, and with quadrant photodiode shutters on given bench
    either open or closed. }
  \label{tab:fr}
\end{table}

Using equation~\eref{eq:P_B8_RIN} we fit the scattering fraction $f_r$ to
match the RIN of $P_\text{B8}$. In
addition, the second order scattering fraction $f_{2r}$ is clearly
visible, which correspond to light that makes two round trips between
the mirror and the bench. An example is shown on
figure~\ref{fig:fringeWrap}, and the fit parameters for all four
measurements are shown in table~\ref{tab:fr}.

Note that closing the quadrant shutters reduces the SWEB scattering by
almost an order of magnitude from $f_r = 4.0 \times 10^{-8}$ to
$f_r = 6.0 \times 10^{-9}$. When the shutters are closed the beam is
dumped on anti-reflective coated black glass on the bench which has
low scattering, which explains the reduction in scattering. However,
for SNEB the scattering increases from $f_r = 2.5 \times 10^{-8}$ to
$f_r = 3.5 \times 10^{-8}$. This is due to the lack of these beam
dumps on SNEB, the beam is sent instead on the vacuum chamber wall
that appears to have a scattering similar in magnitude to the back
scatter from the quadrant photodiode sensor.

\section{Subtraction of scattered light}
\label{sec:subtract}

\begin{figure}
  \centering
  \includegraphics[width=\textwidth]{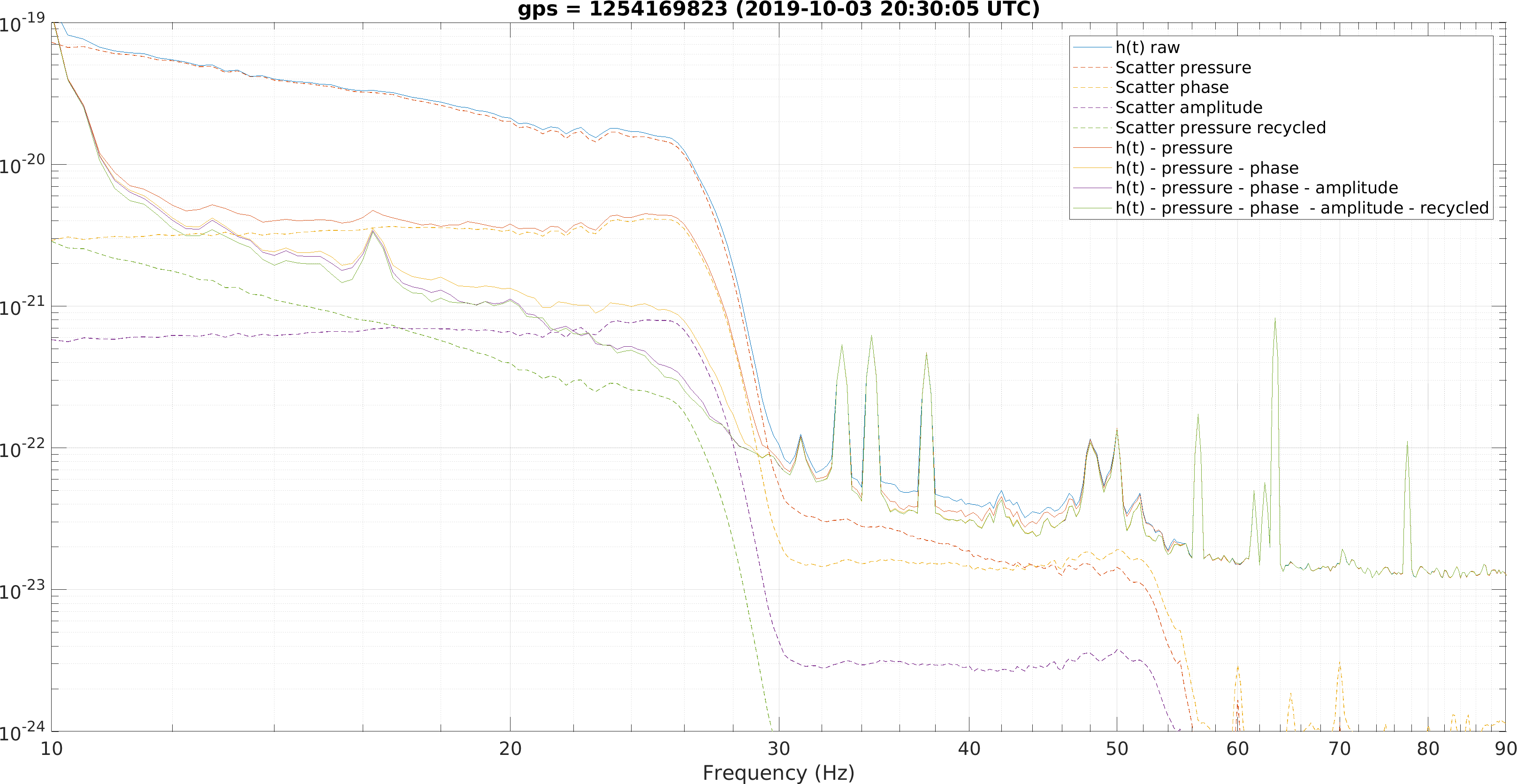}
  \caption{Spectrum of strain data during an intentional motion of
    SWEB. Shown are the original strain data, the projections of the
    radiation pressure, phase and amplitude couplings, alongside the
    successive subtraction in time domain of each of these contributions.}
  \label{fig:subtract}
\end{figure}

The RIN of $P_\text{B8}$ give us direct access to the coherent sum of
$\cos\phi_\text{sc}$ and  $\cos2\phi_\text{sc}$ with the appropriate scattering
fraction coefficient. In addition, the $\sin\phi_\text{sc}$ term which couples
through \eref{eq:sc_phase} can be reconstructed from the Fourier
decomposition of $\cos\phi_\text{sc}$ and the bench speed as follows
\begin{eqnarray}
  \cos \phi_\text{sc}(t) &=& \int_{-\infty}^{\infty} \hat{\phi}_\text{sc}(\omega) e^{i \omega t} \der \omega  \\
  \sin \phi_\text{sc}(t) &=& i\mathop{\rm sign}  \left[\frac{\der \phi_\text{sc}(t)}{\der t}\right]  \int_{0}^{\infty}\left[
  \hat{\phi}_\text{sc}(-\omega) e^{-i \omega t} -
  \hat{\phi}_\text{sc}(\omega) e^{i \omega t} \right]\der \omega.
\label{eq:sin_phi}
\end{eqnarray}
We will not attempt a formal derivation of that relation, instead we show
that it is able to coherently subtract the observed scattering
noise.

In practice we use the local controls to obtain the bench speed in
$\rm\mu m/s$, and use the $\tanh$ function instead of
$\mathop{\rm sign}$ to avoid a discontinuity when the speed is
low. For low speeds the exact value of $\sin \phi_\text{sc}(t)$ does not
matter as its contribution will remain below 10\,Hz.

We have verified that this reconstruction method of $\sin
\phi_\text{sc}(t)$
does not introduce any bias in amplitude or phase
using surrogate data as follows. We
create a bench motion time series by filtering white Gaussian noise
with a resonant pole at 0.1\,Hz with a quality factor of 30. This
yields a simulation of an imperfect intentional motion of the
bench. We add a white noise 3 orders of magnitude lower than the
cosine of the bench phase to simulate the sensing noise of the B8
photodiode. Using the method above we reconstruct the sine of the
bench phase and compare it to the direct computation. The obtained
reconstruction errors are lower than 0.1\% in the 10-30\,Hz band.

We then apply this $\sin \phi_\text{sc}(t)$ reconstruction method to the real detector data
described in section~\ref{sec:measurement}. 
We  use the transfer functions given by equations
(\ref{eq:sc_phase},\ref{eq:sc_amplitude},\ref{eq:sc_pressure}) to subtract
the scattering noise measured by $P_\text{B8}$ and the reconstructed
phase quadrature during a time of intentional motion of SWEB. The
result is shown on figure~\ref{fig:subtract} and demonstrate up to a
factor 40 reduction in scattered light noise.
Also the second order scattering between 30\,Hz and 55\,Hz is
correctly removed. To be able to fit these theoretical transfer
functions to measurements, we diagonalize the measured transfer
functions using the cross correlation matrix between $\cos
\phi_\text{sc}$ and $\sin \phi_\text{sc}$, as performed in
gravitational wave strain noise subtractions \cite{Davis2019}.

\begin{table}
  \centering
  \begin{tabular}{l|ccccc}
    & & & & contrast &  \\
    &$T_\text{WE}$ & $T_\text{NE}$& $|E_0|^2$ & defect & $\rho$  \\
    &(ppm) &  (ppm)& (kW) & ($\mu$W) & (\%)\\
    \hline
    SWEB, open & $4.34 \pm 0.04$ & -- &$92.1\pm0.7$ & $95\pm8$
                                                               & $1.41
                                                                 \pm
                                                                 0.06$\\
    SWEB, closed & $4.36 \pm 0.17$ & -- &$92.1\pm5.1$ & $82\pm31$
                                                               & $1.73 \pm 0.10$\\
    SNEB, open & -- & $4.45 \pm 0.08$ & $91.6\pm1.8$ & $132\pm19$
                                                               & $1.05 \pm 0.17$\\
    SNEB, closed & -- & $4.46 \pm 0.05$ & $91.3\pm1.2$ & $180\pm12$
                                                               & $1.26 \pm 0.04$\\
    \hline
    expected & & & & & \\
    \cite{WEtransmission,NEtransmission,contrastDefectO3a}& $4.3 \pm 0.2$ & $4.4 \pm
                                                    0.1$
        & $90\pm5$ & $120\pm20$ & $\sim
                                                                 1$
     
  \end{tabular}
  \caption{Fitted interferometer parameters for four different
    measurements with intentional large amplitude motion of either
    SWEB or SNEB, and with quadrant photodiodes shutters on given bench
    either open or closed. For comparison the expected value of these
    parameters from other measurements is also shown.}
  \label{tab:fits}
\end{table}

In total four measurements of 3 minutes in duration have been
performed. Two with an intentional motion of SWEB and two with an
intentional motion of SNEB. In each case one of the measurement was
with the quadrants shutters open and one with the quadrants shutters
closed.

To achieve this high subtraction efficiency we fitted the
interferometer parameters as listed in table~\ref{tab:fits}.  The
measurement errors were estimated by splitting the 3 minute of
available data into 6 blocks of 30 seconds, and repeating the complete
analysis and fit separately on each of them. However these do not
include systematic errors for example due to the frequency dependent
response of the photodiodes, which could add $\sim$2\% errors on the
leading parameters of $T_\text{WE}$, $T_\text{NE}$ and $|E_0|^2$, and potentially
larger errors in the sub-dominant parameters. This is a likely
explanation for the inconsistency in the measured contrast defect and
arm power asymmetry $\rho$ between the SNEB and SWEB measurements.

The model and method described above have not been used to subtract
scattered light noise from O3 online strain data used for gravitational wave
analysis. Instead a simpler model independent approach has been used
in the reconstruction process of the strain data to subtract several
different noises~\cite{O2_calibration}. This was performed by measuring the
transfer function between auxiliary channels and the strain data over
a 500\,s long strech of data, and using these transfer functions on
the following 500\,s of data to subtract the noise from these
auxiliary channels. The measured power $P_\text{B8}$ and $P_\text{B7}$
were among the auxiliary channels to subtract noise, which allowed to
subtract the combined contribution of radiation pressure, recycled
radiation pressure and amplitude coupling. $P_\text{B8}$ was more
critical as the SWEB suspension suffered from a reduced isolation
leading to larger motion during bad weather times.

\begin{figure}
  \centering
  \includegraphics[width=\textwidth]{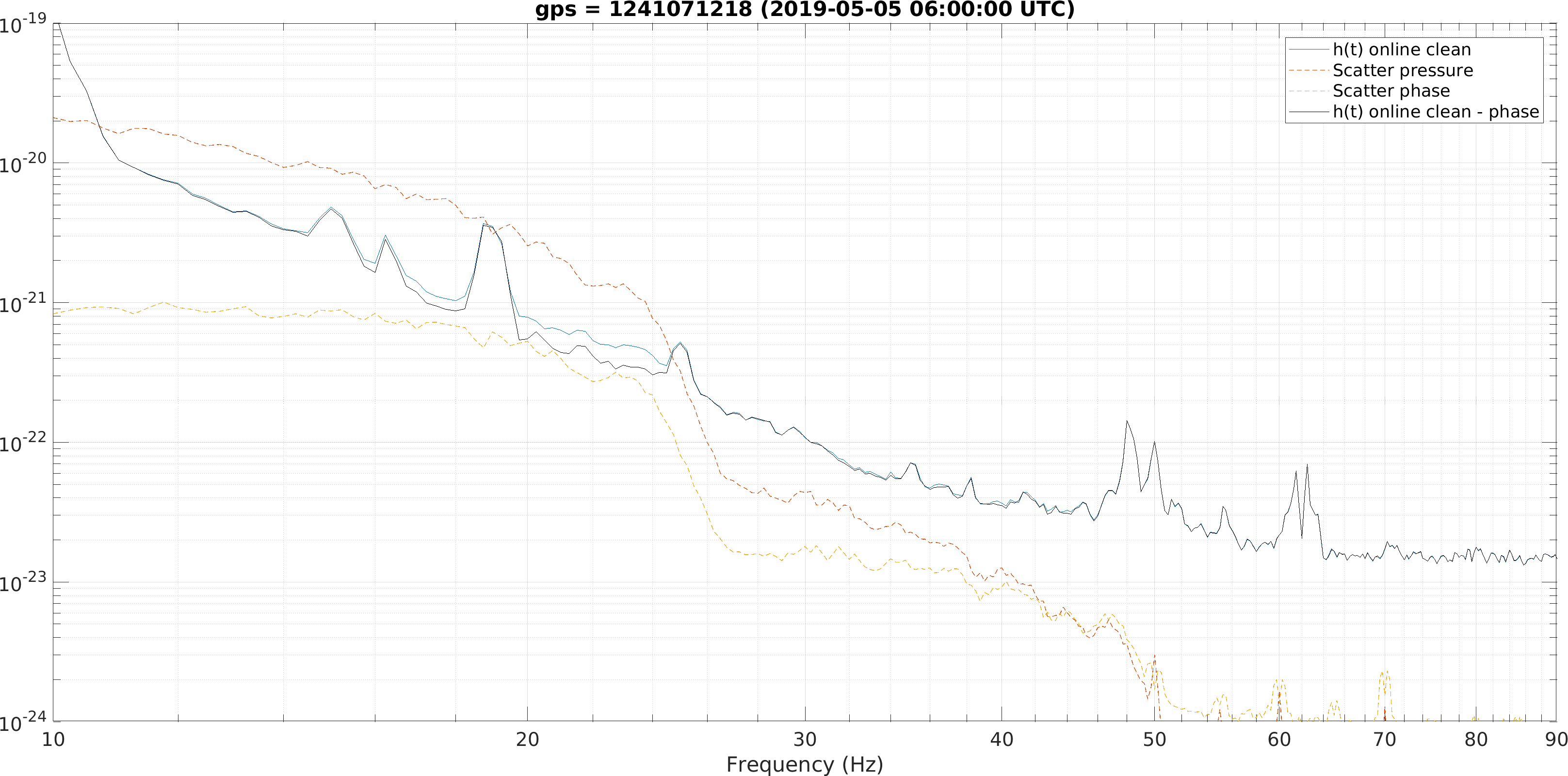}
  \caption{Spectrum of strain data during bad weather. Shown are the
    strain data after online noise cleaning, the projections of the
    radiation pressure and phase couplings of SWEB motion, and the
    effect of subtracting the phase coupling from the already cleaned
    data.}
  \label{fig:subtractWeather}
\end{figure}

On figure~\ref{fig:subtractWeather} we show that this noise
subtraction was effective at removing the radiation pressure
contribution of scattered light from SWEB that is coherent with
$P_\text{B8}$, leading to sensitivity improvement of up to a factor
4. However that method was not able to remove the phase coupling path which
is not coherent with B8. Using equation~\eref{eq:sin_phi} to reconstruct the
phase coupling we are able to reduce the noise by a further 30\% at
20\,Hz.

The strain data after subtraction are incoherent with the two
subtracted components (proportional to $\cos
\phi_\text{sc}$ and $\sin \phi_\text{sc}$), hence the residual strain
data noise adds in quadrature with these components. Given the overall
factor 5 decrease in noise at 20\,Hz, it means that the scattered
light noise measured by $P_\text{B8}$ has been removed with $\sim98\%$
efficiency, consistent with the factor 40 reduction in noise observed
during intentional motion of the benches. 

\section{Interferometer absolute calibration}
\label{sec:calibration}
The measurement of $T_\text{WE}$ and $T_\text{NE}$ described above are
derived from equation~\eref{eq:sc_phase} and assumed that the
calibration of the strain data was accurate. However the transmission
of the end mirrors have also been measured at LMA before installation~\cite{WEtransmission,NEtransmission}
($T^\text{LMA}_\text{WE} = 4.3 \pm 0.2 \times 10^{-6}$ and
$T^\text{LMA}_\text{NE} = 4.4 \pm 0.1 \times 10^{-6}$). Comparing the two
measurement of the mirror transmission yields a measurement of the
accuracy of the strain data calibration.

Indeed, combining equations~\eref{eq:sc_phase} and \eref{eq:P_B8_RIN}  we obtain
that scattered light introduces an absolutely calibrated signal 
\begin{equation}
  h_\text{signal} = \frac{1}{L}  T_\text{WE}
\frac{\lambda}{4 \pi} \frac{1}{2} \mathcal{TF}\left[\frac{\delta P_\text{B8}}{P_\text{B8}}\right],
\end{equation}
where $\mathcal{TF}$ is the transform from to $\cos
\phi_\text{sc}$ to $\sin \phi_\text{sc}$ given in
equation~\eref{eq:sin_phi}. 
The arm length $L$ and the
laser wavelength $\lambda$ are known with precision better than
0.01\%, which introduces a negligible error. We also assume that the
photodiode frequency response is corrected to be flat between DC and
the measurement band of 10-30\,Hz.
In this case the ratio of the
calibrated scatter light signal with the calibrated strain data after reconstruction
$h_\text{rec}$ is directly equal to the ratio of the two estimates of
the WE mirror transmissions
\begin{equation}
  \frac{h_\text{rec}}{h_\text{signal}} =
  \frac{T_\text{WE}}{T^\text{LMA}_\text{WE}} =
  \frac{4.34 \pm 0.04 \times 10^{-6}}{4.3 \pm 0.2 \times 10^{-6}} =
  1.009 \pm 0.05.
\end{equation}
Analogously for the NE mirror transmission estimates yields
$\frac{h_\text{rec}}{h_\text{signal}} = 1.014 \pm 0.025$. The dominant source of error
in these measurements comes from the 2-5\% uncertainty in the end
mirror coating transmission, the statistical error is of the order of
1\% and could be reduced by longer measurements.

The current calibration of LIGO and Virgo is performed using a \emph{photon
calibrator} \cite{LIGO_Pcal,Virgo_Pcal}, an auxiliary laser that
pushes on the mirrors through radiation pressure. A fundamental issue
of that method is the absolute calibration of the laser power of that
auxiliary laser, as the references standards in different countries
are in disagreement by several percent. For O3 this has been addressed by inter-calibrating
the LIGO and Virgo power references, which removes a calibration bias
between the instruments but leaves the possibility of an absolute bias
of the calibration of the gravitational wave detector network. The
method described above could allow an absolute calibration of the
detectors which do not rely on these power reference
standards. Instead the method relies on precise measurement of the
transmission of the end mirrors before their installation, a relative
power measurement that has been performed with a few percent
precision. However, in principle that precision could be significantly
improved by developing the corresponding metrology.

\section{Conclusions}
\label{sec:conclusion}

We have shown how both quadratures of the scattered light from
suspended end benches can be reconstructed from the signal of the
photodiode located on that bench and from the information on the sign
of the bench displacement speed. We derive a model of the main
coupling mechanism for scattered light from these benches to the
detector sensitivity, which allows to fit real data and subtract the
scattered light noise contribution by up to a factor 40.

Moreover, the fitted interferometer parameters demonstrate how
scattered light injections can be used to characterize the
interferometer, as they provide self calibration injections of light
field fluctuations directly into each arm cavity. In particular they
are able to measure the power circulating in the interferometer and
the interferometer contrast defect.

We have also shown that scattered light noise can be used to
accurately calibrate the absolute response of the detector. This
method is completely independent of previously proposed methods such
as the ``free Michelson'' method~\cite{O2_calibration}, the photon calibrator or \emph{newtonian
  calibrator} \cite{Estevez2018}. We stress that measurement described
above was opportunistic and performed for a different
purpose. Dedicated measurements would yield more robust results.
For instance by performing scattered light injection over a longer
time with a faster motion to cover frequencies up to at least 100\,Hz,
which more clearly separates the different coupling mechanism that are
dominant at different frequencies.

\ack

The authors gratefully acknowledge the Italian Istituto Nazionale di
Fisica Nucleare (INFN), the French Centre National de la Recherche
Scientifique (CNRS) and the Netherlands Organization for Scientific
Research, for the construction and operation of the Virgo detector and
the creation and support of the EGO consortium.  The authors also
gratefully acknowledge research support from these agencies as well as
by the Spanish Agencia Estatal de Investigaci\'on, the Consellera
d'Innovaci\'o, Universitats, Ci\`encia i Societat Digital de la
Generalitat Valenciana and the CERCA Programme Generalitat de
Catalunya, Spain, the National Science Centre of Poland and the
Foundation for Polish Science (FNP), the European Commission, the
Hungarian Scientific Research Fund (OTKA), the French Lyon Institute
of Origins (LIO), the Belgian Fonds de la Recherche Scientifique
(FRS-FNRS), Actions de Recherche Concertées (ARC) and Fonds
Wetenschappelijk Onderzoek – Vlaanderen (FWO), Belgium.
The authors gratefully acknowledge the support of the NSF, STFC, INFN,
CNRS and Nikhef for provision of computational resources.

\section*{References}

\bibliographystyle{unsrt}

\end{document}